\documentclass[11pt]{article}%
\usepackage[slantedGreek]{mathpazo}
\usepackage{amsmath,amssymb,amsthm,amsfonts}
\usepackage[left=0.9in,top=0.9in,right=0.9in,bottom=0.9in]{geometry}
\usepackage{graphicx}
\usepackage{subcaption}
\usepackage{suffix}
\usepackage{color}
\usepackage{setspace}

\usepackage{datetime}

\RequirePackage[colorlinks,citecolor=blue,urlcolor=blue]{hyperref}

\newdateformat{monthyear}{\monthname[\THEMONTH] \THEYEAR}

\allowdisplaybreaks

\graphicspath{{../../figures/}{../figures/}{./figures/}{./}}

\vspace{-0.7in}
\title{From Least Squares to Signal Processing and Particle Filtering}

\author{
  Nozer D. Singpurwalla\\
    \textit{City University of Hong Kong}
  \footnote{email: nsingpur@um.cityu.edu.hk }
  \and
  Nicholas G. Polson \\
    \textit{University of Chicago}
  \footnote{email: ngp@chicagobooth.edu}
  \and
  Refik Soyer \\
  \textit{George Washington University}
    \footnote{email: soyer@gwu.edu}
}

\begin{document}
\maketitle

\begin{abstract}
  \noindent 
 De facto, signal processing is the interpolation  and
extrapolation of a sequence of observations viewed as a
realization of a stochastic process. Its role in applied statistics ranges from scenarios in forecasting and time series analysis, to image reconstruction, machine learning, and the degradation modeling for reliability assessment. This topic, which has an old and honourable
history dating back to the times of Gauss and Legendre, should therefore be of interest to readers of \textit{Technometrics}. A general solution to
the problem of filtering and prediction entails some formidable mathematics.
Efforts to circumvent the mathematics has resulted in the need for
introducing more explicit descriptions of the underlying process.
One such example, and a noteworthy one, is the Kalman Filter Model, which is a special case of state space models or what
statisticians refer to as  Dynamic Linear Models. Implementing the Kalman
Filter Model in the era of ``big and high velocity non-Gaussian data'' can pose computational challenges with
respect to efficiency and timeliness. Particle filtering is a way to ease
such computational burdens. The purpose of this paper is to trace the
historical evolution of this development from its inception to its current
state, with an expository focus on two versions of the particle filter, namely, the propagate first-update next and the update first-propagate next version.

By way of going beyond a pure review, this paper also makes transparent the importance and the role of a less recognized principle, namely, the \textit{principle of conditionalization}, in filtering and prediction based on Bayesian methods. Furthermore, the paper also articulates the philosophical underpinnings of the filtering and prediction set-up, a matter that needs to be made explicit, and  Yule's decomposition of a random variable in terms of a sequence of innovations. 
\vspace{0.1in}

\noindent\textit{Keywords: Dynamic linear models, Filtering likelihood, Genetic Monte Carlo, Kalman filter, Machine learning, Prediction, Smoothing likelihood, State space models, Reliability,  Time series analysis}. 
\end{abstract}

\newpage

\setstretch{1.5}

\section{Antecedents to Signal Processing and Smoothing}

It is fair to state that the genesis of signal processing is the work in
1795 of an 18 year-old Gauss on the method of least squares. The motivation for Gauss' work was astronomical studies on planet
motion using telescopic data. Though this work was formally published only in
1809, Gauss laid out a general paradigm for all that has followed. In
particular, he recognized that observations are merely approximations to the
truth, that such errors of measurement call for more observations than the
minimum required to determine the unknowns, that one needs to invoke dynamic
models (such as Kepler's laws of motion) for estimating the unknowns, and
that a minimization of a function of the residuals leads to their most accurate assessment.
More importantly, Gauss also addressed the matter of suitable combination of
observations that will provide the most accurate estimates. The above in turn
gave birth to the design of filters as a linear or non-linear combination of
observables. On page 269 of his \textit{Theoria Motus Corporum Coelestium} (1809),
Gauss predicted that his principle of least squares could spawn countless
methods and devices by means of which numerical calculations can be
expeditiously rendered. This opened the door for approaches like the Kalman
Filter to thrive and to survive. Thus, in effect, the Kalman Filter is an
efficient computational device to solve the least squares problem, and the
particle filter enhances the efficiency of such computational algorithms by speeding them
up and by allowing them to be applied in non-Gaussian contexts. But the journey from the ideas of Gauss to the currently popular particle
filtering took over 200 years to complete, with the likes of Kolmogorov,
Wiener, Bode, and Shannon in the driver's seat. Indeed, as suggested by a referee, a more appropriate title of this paper should have been "From Least Squares to Particle Filtering," but doing so could have detracted the attention of control theorists and signal processors who may view the topic of least squares as being predominantly statistical in nature.

It was almost 135 years after Gauss enunciated the key principles of
estimation that Kolmogorov in 1939 provided his solution to the problem of
interpolation and extrapolation with minimal assumptions. Specifically,
Kolmogorov assumed that the underlying stochastic process is discrete in
time, is stationary, and has finite second moments. This set the stage for
all that is to follow, including Kolmogorov's 1940 paper which embeds the
problem scenario in a Hilbert space and reduces his results of 1939 as a
special case. Kolmogorov's 1940 paper is a tour de force in elegance and
mathematical genius comprising of just 9 pages. One may wonder as to why
problems of interpolation and extrapolation continue to persist despite
its closure brought about by the above works. However, an examination of
Kolmogorov's results, overviewed in Section 2 of this paper, reveals their
formidable nature, and the difficulty in putting them to work.

At about the same time as Kolmogorov, circa 1942, Wiener working on the World
War II problem of where to aim anti-aircraft guns at dodging airplanes
arrived upon the continuous time formulation of the interpolation and
extrapolation problem, now known as "signal processing." Here,
interpolation got labeled as ``filtering'' (or smoothing) and extrapolation
as ''prediction.`` Wiener's work, presumed to be done independently of that
by Kolmogorov, was for the National Defense Research Council, and
remained classified until 1949, when it was reprinted as a book \lbrack
Wiener (1949)\rbrack. Like Kolmogorov's work, Wiener's work was also
mathematically formidable involving the notoriously famous Wiener-Hopf
equation. In Section 3 we give an outline of Wiener's work leading up to the
above mentioned equation (which does not arise in the discrete case of a signal plus noise model). A noteworthy feature of Section 3, is Section 3.1, wherein the philosophical underpinnings of the Kolmogorov-Wiener setup are articulated, especially as they relate to the spirit and the excitement of the early 1920's and 1940's, namely quantum theory. It behooves those interested in filtering, extrapolation and machine learning, to be cognizant of what is it that spawned the models they engage with. 

The material of Sections 2 and 3 gives the reader an appreciation for
the need to develop efficient computational devices like the Kalman filter
and the particle filter, which can now be seen as a computational device
overlaid on another computational device in order to generalize and speed up the former. The remainder of this paper is
organized as follows: Section 4 pertains to the genesis of the state space
models via the works of Bode and Shannon (1950), and of Zadeh and Ragazzini
(1950), based on electrical circuit theory. The important role played by these works in the development of the statistical theory of dynamic models seems to be unappreciated. Section 5 pertains to the Kalman
Filter Model as prescribed by Kalman in 1960, and its (relatively less appreciated) relationship to Yule's
random innovations and the Box-Jenkins approach it spawned, and to Doob's conditional expectation. Section 6 continues
with the theme of Section 5 by providing an outline of the Bayesian prior to
posterior iteration which is the essence of Kalman's filtering algorithm.
Whereas the material of Section 6 is well known (to most statisticians and
signal processors), it is presented here to set the stage for the material of
Section 7 on particle filtering whose exposition, albeit cursory, is a part of the objectives
of this paper. Section 6 also highlights the routinely invoked, but less recognized, \textit{principle of conditionalization}, implicit to Kalman filtering. 
Section 8 concludes the paper with some conjectures about the path forward.

The value of this paper rests on its expository character, vis a vis tracing the historical development from signal processing to particle filtering, articulating the principle of conditionalization, the philosophical underpinnings of the Kolmogorov-Wiener setup and the relationship between the Kalman filter model and Yule's (1927) statistically motivated notion of random innovations, also known as "random shocks".

\section{Kolmogorov's Interpolation and Extrapolation of a Sequence}

Specifically, for a random variable $X(t)$, with $t$ an integer and
$-\infty<t<+\infty$, suppose that $E[X^2(t)]<\infty$, and that the sequence $\{X(t);-\infty<t<+\infty\}$ is
stationary. Without loss of generality set $E[X(t)]=0$, and note that
$B(k)=E[X(t+k)X(t)]=B(-k)$, the autocorrelation at lag $k$, will not depend
on $t$, for any integer $k\geq 0$. The problem of linear extrapolation is to
select for any $n>0$ and $m>0$, real coefficients $a_i$, for which
\begin{equation*}
L=a_1X(t-1)+a_2X(t-2)+\cdots+a_nX(t-n)
\end{equation*}
gives the closest approximation to $X(t+m)$. As a measure of accuracy of this
approximation, Kolmogorov (1939) leans on the Gaussian paradigm of minimizing the error sum of squares and considers $\sigma^2=E[(X(t+m)-L)^2]$
to seek values of $a_i$ for which $\sigma^2$ is a minimum. 
If this minimum
value is denoted by $\sigma_\mathcal{E}^2(n,m)$, then Kolmogorov shows that $\sigma_\mathcal{E}^2(n,m)$ has a limit, and he uses this limit to find the minimizing $a_i$'s.

For the interpolation part, the estimation of $X(t)$ using
$X(t\pm 1),\,X(t\pm 2),\cdots,X(t\pm n)$ is considered, so that if
\begin{equation*}
Q=a_1X(t+1)+\cdots+a_nX(t+n)+a_{-1}X(t-1)+\cdots+a_{-n}X(t-n)\text{,}
\end{equation*}
then the problem boils down to minimizing $\sigma^2=E[(X(t)-Q)^2]$. If
$\sigma_\mathcal{I}^2(n)$ denotes this minimum, then
$\sigma_\mathcal{I}^2(n)$ cannot increase in $n$ and so its limit, $\sigma_\mathcal{I}^2$,   exists, and Kolmogorov finds this limit. In both of the above cases, Kolmogorov uses  formidable mathematics
pertaining to the spectral theory of stationary processes. This underscores the point made before that interpolation and extrapolation are difficult tasks.

\section{Wiener's Theory: The Birth of Statistical Signal Processing}

Whereas Kolmogorov's approach is cast in the language of probability, Wiener [cf. Wiener (1949)]
casts his in the language of communications theory (and hence signal
processing). More significantly, Wiener considers the continuous case, and
endows the set-up with additional structure than that of Kolmogorov's.
Specifically, an observable random sequence $y(t)\,$is decomposed as the sum
of a random signal $s(t)$ and perturbing noise $n(t)$, unrelated with $s(t)$;
that is, $y(t)=s(t)+n(t)$. It is desired to operate on the $y(t)$'s in such a way
so as to obtain, as well as is possible, the signal $s(t)$. The act of
operating on the $y(t)$'s, became known as filtering, and a filter is a
precise specification of the operation on $y(t)$. Wiener also considers the
combining of a filtering operation with prediction. That is, operating on
$y(t)\,$to obtain a good approximation to $s(t+\alpha)$, for some $\alpha>$or
$<0$.

Underlying Wiener's approach are three assumptions. These are: that the
stochastic processes generating the signal $s(t)$ and the noise $n(t)$ are
stationary with finite second moments, that $s(t)$ is independent of
$n(t)$, that the criterion for the error of approximation is mean square
discrepancy, and that the operator on $y(t)$ for filtering and prediction is
to be linear on the available information and be implementable (i.e. a computable function of the observed data assuming the availability of the data). In the language
of communication theory, the filter is to be linear (in the observed data) and physically
realizable (to be explained later). The available information is the past history of the perturbed
signal $y(t)$. The assumptions of Wiener parallel those of
Kolmogorov; the key differences between the two being a discrete $t$ versus a
continuous $t$, and a decomposition of the observable $y(t)$ into the form of a signal $s(t)$ and a
noise $n(t)$. Even so, the probabilistic architecture underlying the two set-ups is
identical.

\subsection{Philosophical Underpinnings of the Kolmogorov-Wiener Setup}

Predicting the future behavior of a signal based on a perturbed version of
its present and past history is grounded in philosophical issues pertaining to causality,
induction, and the nature of physical law. In general, prediction is based on the inductive
premise that the observed patterns of the past will continue to be so
in the future. This in turn is an assumption which implies that the past is the cause
of the future. An assumption of causality like this one cannot be deduced mathematically.
It can not be established empirically either, because empirical verification using statistical techniques entails the null hypothesis that the cause-effect relationship is true, and then an investigation to see if the evidence causes a rejection of the hypothesis. Indeed, the
notion that the past is a guide to the future is a central postulate of all
the empirical sciences. Classical physics attempted to describe the physical
world via a set of (deterministic) causal laws whose role was to relate the
past to the future. Examples are: Newton's Laws, Kepler's Law, Ohm's Law,
and so on. Quantum physics denied such laws, and claimed them untenable for the
microscopic world. Quantum physics claims that on an atomic scale, the laws of physics are
only statistical, and that the only meaningful predictions are statistical.

The Kolmogorov-Wiener set-up adheres to the above quantum physics based view that
all predictions are statistical, and so is the causal relationship between
the past and the future. This viewpoint is asserted via two assumptions:
stationarity, and the existence of second moments (i.e. correlations). Prediction is based on the existence 
of a correlation between the future
values of the signal and the past values of the observables, and correlation
is indeed the manifestation of a statistical relationship. Kolmogorov's requirement
of a finite second moment of $X(t)$ is an assertion of the above thesis.
Furthermore, the Kolmogorov-Wiener requirement that the filter be linear,
is tantamount to the feature that the only type of relationship that needs to be considered, is a linear, and this manifests itself as a correlation.

To summarize, the routinely invoked Kolmogorov-Wiener assumptions of stationarity, finite
second moments, and filter linearity are dictated by the philosophical
considerations underlying causality and predictivity. Making this matter explicit is a feature of this paper, and one which enhances its expository character; also see Cox (1992).

\subsection{Filtering, Prediction and the Wiener-Hopf Equation}

We start with  the Wiener-Hopf equation, and trace the 
 steps that lead to it.

Suppose that $\varphi(x)\,$is an unknown function of $x$, $0\leq x<\infty$,
and $K(\cdot)$ and $f(\cdot)$ are known functions with $K(\cdot)$ being
monotone. Suppose that for $x>0$,
\begin{equation*}
\varphi(x)=-\int_0^\infty\varphi(y)K(x-y)\,dy+f(x)\text{,}\tag*{(3.1)}
\end{equation*}
and it is required that the solution to this equation be of the form
\begin{equation*}
\varphi(x)\leq c<\infty\text{, where lim
}_{x \rightarrow \infty}\varphi(x)=c.
\end{equation*}
(3.1) is the Wiener-Hopf equation, with equivalent representation:
\begin{equation*}
\varphi(x)=-\int_{-\infty}^\infty\varphi(x-y)\,dK(y)+f(x)\text{,}\,0\leq
x<\infty.\tag*{(3.2)}
\end{equation*}
(3.2) has been notoriously difficult to solve in general (for processes whose spectral densities are not rational), and 
attempts to get computationally efficient solutions have lead to approaches like the Kalman
Filter. This is the topic of the next section. For now we outline the steps
which led to it. The material here is abstracted from Davenport and Root
(1958), p. 219.

A filter, $h(t)$, is a weighting function operating on
$y(t)$ to give 
\begin{equation*}
\int_{-\infty}^\infty h(t-\tau)\,y(\tau)\,d\tau=\int_{-\infty}^\infty
h(\tau)\,y(t-\tau)\,d\tau\text{,}
\end{equation*}
with the requirement of \textit{physical realizability}, which means that
$h(t)=0$, for $t<0$. One approach towards advocating the efficacy of the filter is to require that $h(\cdot)$ be
chosen so that $\mathcal{E}$, the expected mean square, is minimized. Here, $s$ is the process to be estimated, and $y$ is the observable process:
\begin{equation*}
\mathcal{E}=E\bigl\{\bigl[s(t+\alpha)-\int_{-\infty}^\infty
h(\tau)\,y(t-\tau)\,d\tau\bigr]^2\bigr\}.\tag*{(3.3)}
\end{equation*}

Since $s(t)$ and $n(t)$ are stationary, independent, and have finite
second moments, their auto and cross-correlations exist, and are
time invariant. Consequently, 
\begin{equation*}
\mathcal{E}=E\bigl[s^2(t+\alpha)\bigr]-2\int_{-\infty}^\infty
h(\tau)\,E\bigl[s(t+\alpha)\,y(t-\tau)\bigr]d\tau
\end{equation*}
\begin{equation*}
+\,\int_{-\infty}^\infty\int_{-\infty}^\infty
h(\tau)h(\mu)E\bigl[y(t-\tau)\,y(t-\mu)\bigr]d\tau\,d\mu
\end{equation*}
\begin{equation*}
=B_s(0)-2\int_{-\infty}^\infty
h(\tau)\,B_{sy}(\alpha+\tau)\,d\tau+\int_{-\infty}^\infty\int_{-\infty}^%
\infty h(\tau)h(\mu)\,B_y(\tau-\mu)d\tau\,d\mu\text{;}
\end{equation*}
where $B_s(k)$ is the autocorrelation at $k$ of the signal process,
$B_y(k)$ the autocorrelation of the observable process, and $B_{sy}(k)$ the
cross-correlation at $k$ of these processes.

It is shown \lbrack Davenport
and Root (1958), p. 223-224\rbrack\ that a necessary and sufficient condition
$h(t)$ must satisfy for $\mathcal{E}$ to be a minimum is
\begin{equation*}
B_{sy}(\tau+\alpha)=\int_0^\infty h(\mu)\,B_y(\tau-\mu)\,d\mu\text{,
}\tau\geq 0.\tag*{(3.4)}
\end{equation*}
 
The above  is an integral equation which relates a cross-correlation
with an autocorrelation, and in the context of the philosophical material of Section 3.1, can be
interpreted as a statistical law. The solution to (3.4) will yield
an optimum smoothing and prediction filter, and the challenge here has been to find
a solution. It has been shown that an exact solution to a realizable filter is based on the requirement that $S_y(f)$, the Fourier transform of $B_y(\tau)$, be rational (so that it can be easily factored), and the solution is expressed in terms of the factors of $S_y(f)$ and $S_{ys}(f)$, the cross spectral density of the $y(t)$ and $s(t)$.
The solution therefore has been challenging to obtain, and this has spawned derivations alternate to the above, the pioneering ones 
being those by Bode and Shannon (April 1950) and by Zadeh and Ragazzini (July
1950). 
The underlying concept behind both the above approaches was to give a more
explicit description of the signal by introducing an additional filter,
called the ``shaping filter.'' Whereas the statistics community (and possibly also the machine learning community) is well aware of the Kalman filter model, the pioneering works of Bode and Shannon and of Zadeh and Ragazzini which gave birth to \textit{state-space models}, of which the Kalman filter is a special case, appears to be less recognized by the above community (communities). A purpose of this paper is to correct this possible skewness and highlight these overlooked historical footprints.

\section{Precursors to Kalman Filtering: The Shaping and Matched Filters}

The notion of introducing a shaping filter first appeared in Bode and Shannon
(1950) whose aim was to develop a simplified approach for smoothing and
filtering under Wiener's set up. The underpinnings of their approach, (which is a simple representation of white noise), was
based on circuit design, and their discussion was cast in the language of
communications theory entailing the notions of impulses and responses. Based
on the first several readings of the Bode-Shannon paper, it is difficult to
see as to how the material therein gave birth to the Kalman Filter Model and
the other dynamic linear models which followed. But once the fog of
terminology is cleared, the ideas become more transparent.

The starting point of the Bode-Shannon approach is a decomposition of a response $s(t)$, not necessarily the $s(t)$ of $y(t)=s(t)+n(t)$. This entails the introduction of a Shaping Filter, the inputs
to which are a large number of closely spaced short impulses over time; see Figure 1. The Shaping Filter produces a response to each impulse, so that
the response at time $t$ spawned by impulse $i$ is some function $s_i(t)$; see Figure 1.
For a linear filter, the responses add up to produce
$s(t)=\sum_is_i(t)$, the total response of the shaping filter. 

\begin{figure}
\begin{center}
\includegraphics[scale=0.65]{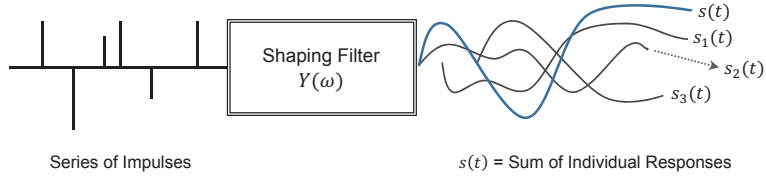}
\end{center}
\label{f1}
\caption{Impulses and Response of the Shaping Filter.}
\end{figure}

The shaping filter is characterized by its response to a unit impulse
impressed on it at time $0$. Thus, for example, if $K(t)$ is the response of
a shaping filter at time $t>0$, to a unit impulse at time $0$, then
$Y(\omega)$, the transfer function of the shaping filter, is the Fourier
transform of $K(t)$; namely, the complex function
\begin{equation*}
Y(\omega)=\int_{-\infty}^\infty K(t)\,e^{-j2\pi\omega t}\,dt.
\end{equation*}
Conversely, $K(t)$ is the inverse transform of $Y(\omega)$; see Figure 2. Thus 
\begin{equation*}
K(t)=\int_{-\infty}^\infty Y(\omega)\,e^{j2\pi\omega t}\,d\omega.
\end{equation*}

\begin{figure}[!ht]
\begin{center}
\includegraphics[scale=0.65]{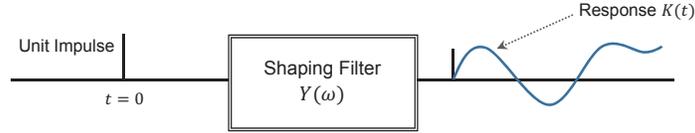}
\end{center}
\label{f2}
\caption{Response of Shaping Filter to a Unit Impulse.}
\end{figure}

Motivated by this line of thinking, the response of the shaping filter
to any continuous input, say $Z(t)$, is obtained by breaking up $Z(t)$ into a
large number of thin vertical slices and regarding each slice as an impulse
of strength $Z(t)dt$. An impulse of strength $Z(t)dt$ impressed on the
shaping filter at time $t$ will produce a response $Z(t)dtK(t_1-t)$ at $t_1$, so that $g(t_1)$, the total response of the filter is:
\begin{equation*}
g(t_1)=\int_{-\infty}^{t_1}Z(t)K(t_1-t)dt\text{, or}
\end{equation*}
\begin{equation*}
=\int_{-\infty}^{t_1}Z(t_1-t)K(\tau)d\tau.
\end{equation*}
If realizability is a requirement, then $K(\tau)=0$, for $\tau<0$.

If the input function $Z(t)$ is deterministic, then so will be its output
$g(t_1)$ for $t_1>0$. In Wiener's set-up, the signal $s(t)$ is assumed to be
a stationary random process. The shaping filter which is presumed to generate
$s(t)$ needs to have inputs that are impulses of random strength. To achieve
the above Bode and Shannon assume that the closely spaced short impulses are
independent, and have a common Gaussian distribution. The responses of these
impulses add up to generate the stochastic process $s(t)$. It may be of
interest to note that the Shaping Filter is merely a conceptual device
introduced by Bode and Shannon to structure the input
signal $s(t)$ of Wiener's set-up. As such, the Shaping Filter need not be
realizable. The genesis of the Shaping Filter lies in circuit theory and
pertains to the effects of a resistor on an electrical input. 

The work of Zadeh and Ragazzini (1950) builds on the Bode-Shannon theme by generalizing it to
assume that the signal $s(t)$ entails two parts, a stochastic process
$\tilde{s}(t)$ on which is superimposed a deterministic part
$N(t)$ which is a polynomial in $t$ of degree $n$, but with coefficients that are
unknown. Furthermore, Zadeh and Ragazzini require that $h(t)$, the weighting
function of the smoothing and prediction filter vanish outside the range
$0\leq t\leq T$, for a specified $T$.

To recap, the effect of the Bode-Shannon, and the Zadeh-Ragazzini work is
to expand the scope of Wiener set up by giving structure to the observed process
via a shaping filter. As shown in Figure 3, the shaping filter (which is
purely conceptual) precedes the desired filtering, and prediction filter, and
this set-up constitutes a foundation for what are known as \textit{state-space models}.

\begin{figure}[!ht]
\begin{center}
\includegraphics[scale=0.65]{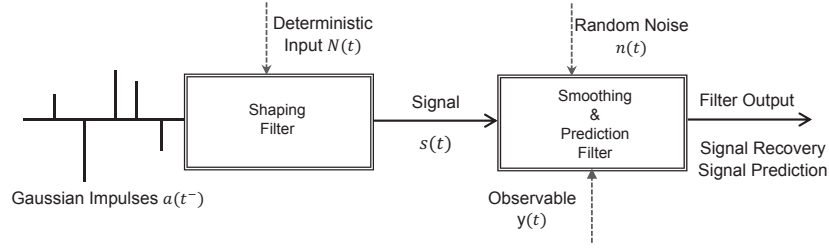}
\end{center}
\label{f3}
\caption{Tandem
Architecture of Shaping and Smoothing Filter.}
\end{figure}

Since the $s(t)$'s share the impulse inputs, denoted by $a(t^{-})$ in Figure 3, they will be
dependent, and as a consequence, so will the $y(t)$'s. This is despite the
fact that $a(t^{-})$'s are independent. It is well known that a
collection of dependent random variables can always be constructed by considering certain functions of 
a collection of independent
variables; see for example, Singpurwalla et al. (2016). A way to mathematically encapsulate the architecture of Figure
3, ignoring the presence of the deterministic function $N(t)$, and
discretizing $t$, as $t=0,1,\ldots$, is to write: 
\begin{equation*}
s(t)=\mathcal{F}[s(t-1)]+a(t)\text{, and }\tag*{(4.1.a)}
\end{equation*}
\begin{equation*}
y(t)=s(t)+n(t).\tag*{(4.1.b)}
\end{equation*}
Here $\mathcal{F}$ is some function of $s(t)$, and the relationships above
constitute the essence of a state space model of which the Kalman Filter Model \lbrack Kalman (1960)\rbrack, with
equation (4.1.a) constituting the dynamic part, is a special case. Linear cases are those in which the relationship between $y(t)$ and $s(t)$ is linear $-$as indicated in (4.1.b)$-$ and so is the relationship between $s(t)$ and $s(t-1)$. Otherwise, the cases are nonlinear.  

Preceding the work of Bode and Shannon (1950), and that of Zadeh and Ragazzini (1950), is the unpublished work of North (1943), and the published work of van Vleck and Middleton (1946) on what is known as "matched filters" \lbrack cf. Turin (1960)\rbrack. Underlying the idea of a matched filter is the requirement that a signal $s(t)$ be a deterministic and of known waveform, as opposed to a stochastic process. When such is the case, the smoothing filter $h(t)$ is easy to specify via an inverse Fourier transform. Such a filter is known as a matched filter because it is matched to $s(t)$, and its virtue is an enhanced ability to detect the presence or the absence of a signal $s(t)$. With $s(t)$ fully specified, the matched filter can be seen as a stepping stone to a structured stochastic process like the Kalman Filter.

There are many scenarios in signal processing wherein matched filters arise naturally \lbrack see Section VI of Turin (1960)\rbrack. They offer potential in non-signal processing applications,  whenever a knowledge of $s(t)$ can be had either via the science of the scenario, or via empirical observations. A well illustrated case in point is the detection of cracks in a material via vibrothermography, discussed in good detail, by Li, Holland, and Meeker (2010). These authors consider the more complex scenario of filtering in three dimensions, and the three-dimensional signal can be specified using heat-dispersion theory, or via an empirical argument.

\section{Relationship to Yule's Innovations and Doob's Conditional Expectations}

Equation (4.1.a) of the Kalman Filter Model has a precedence and a parallel in the manner in which Yule
(1927) conceptualized the autoregressive and the moving average processes of
time series analysis, developed and popularized by Box and Jenkins (1970). Yule proposed 
the notion that a highly dependent series
$s(t)$ is generated by an \textit{innovation series} $a(t)$, where $a(t)$'s are
independent and identically normally distributed with mean $0$ and variance
$\sigma_a^2$. Yule's \textit{causal linear filter} transforms the process $a(t)$ to the process
$s(t)$ via the linear operation
\begin{equation*}
s(t)=\mu+\psi_0a(t)+\psi_1a(t-1)+\psi_2a(t-2)+\cdots\cdots\text{,}
\end{equation*}
where $\mu$, $\psi_0$, and the $\psi_i$'s are unknown constants. Setting
$\mu=0$ and $\psi_0=1$, it can be seen that
\begin{equation*}
s(t)=a(t)+\phi_1s(t-1)+\phi_2s(t-2)+\phi_3s(t-3)+\cdots\cdots, 
\end{equation*}
where the $\phi_i$'s are related to the $\psi_i$'s. Thus $s(t)$ is regressed
on its previous values, and the resulting process is an autoregressive
process. If the coefficients $\psi_i$ are so chosen that $\phi_i=0$ for
$i\geq 2$, then the results is an autoregressive process of order one, which
is equation (4.1.a). Observe the parallel between Yule's construction and the 
Bode-Shannon set-up as encapsulated in Figure 3. Yule's linear filter is
Bode and Shannon's shaping filter; the latter has the advantage of time dependent weights whereas the former does not. With the Kalman Filter Model, we go an
additional step beyond Yule's construction towards a smoothing and prediction
filter. In effect, the $\psi_i$'s of Yule's linear filter capture the essence
of the shaping filter's $Y(\omega)$.

\subsection{Filtering with Conditional Expectations: Martingales}
It has been recognized \lbrack cf. Kalman (1960)\rbrack\ that the Wiener
problem can also be approached from the point of view of conditional
distributions and expectations. This perspective obviates the need to engage
with circuit theory, whitening filters, and the language of signal
processing. All that is needed is a knowledge of probability at the
intermediate level, and a facility with manipulations that are mathematically
cunning. The rewards are plenty, because now one need not be restricted to
linear filters, and more importantly, one can lean on the powerful machinery
of martingales.

We start by focusing on $(y(t)-y(t-1))$ the change experienced by the
observable process $y(t)$, between $(t-1)$ and time $t$; assume for now that
$t$ is discrete. We then ask what is the ``best'' prediction of
$(y(t)-y(t-1))$ ? A meaningful answer [cf. Kailath (1968)], it seems, would be the conditional
expectation
\begin{equation*}
E[y(t)-y(t-1)|y(1),\ldots,y(t-1)]=V(t).\tag*{(5.1)}
\end{equation*}
That is, $V(t)$ is the predicted change in $y(t)$ at time $t$; it is based on
a conditional expectation. Next, one considers the error in predicting $y(t)$
using $V(t)$. That is, the \textit{innovation}
\begin{equation*}
y(t)-E[y(t)|y(1),\ldots,y(t-1)].\tag*{(5.2)}
\end{equation*}
Let $U(t)=\sum_{j=1}^tV(j)$, the sum of the predicted changes, and
\begin{equation*}
M(t)=\sum_{j=1}^t\bigl[y(j)-E[y(j)|y(1),\ldots,y(j-1)]\bigr]\text{,}
\end{equation*}
the sum of prediction errors. It is now easy to see that
\begin{equation*}
y(t)=U(t)+M(t).\tag*{(5.3)}
\end{equation*}

This means that the sum of all changes in the $y(t)$'s, namely $y(t)$ itself,
equals the sum of all the predicted changes $\sum_{j=1}^tV(j)$ in $y(t)$ plus
$M(t)$, the sum of all the predicted errors. To achieve some semblance with
Wiener's set-up, namely that $y(t)=s(t)+n(t)$, we invoke the relationship of equation
(5.3) to write
\begin{equation*}
y(t)-y(t-1)=V(t)+\bigl(M(t)-M(t-1)\bigr).\tag*{(5.4)}
\end{equation*}

More about the quantity $M(t)-M(t-1)$ will be said later, but we first remark
that equation (5.3) is known as \textit{Doob's decomposition} of any observable
process $y(t)$. Simple as it may seem, Doob's decomposition has some
powerful implications, the first of which is that it gives birth to a
martingale process. 

Specifically, it can be verified$-$after some routine algebra$-$that
$E[y(t)|y(1),\ldots,y(t-1)]=M(t-1)$, and this implies that $M(t)$ is a
\textit{martingale with respect to the process} $y(t)$. Furthermore, it can be shown
that
\begin{equation*}
E\bigl[M(t)-M(t-1)\bigr]=0\text{, and that }\tag*{(5.5a)}
\end{equation*}
\begin{equation*}
E\bigl[\bigl(M(t)-M(t-1)\bigr)\bigl(M(t-1)-M(t-2\bigr)\bigr]=0.\tag*{(5.5b)}
\end{equation*}

Thus if the martingale difference $\bigl(M(t)-M(t-1)\bigr)$ of equation (5.4)
can be regarded as an error term, then its essence is that the errors have
zero mean and are uncorrelated (but not necessarily independent). Equation
(5.5b) is the orthogonal increments property of martingales, and is a weakening
of the independent increments property assumed in set-ups like classical
regression.

Equation (5.3) is quite general and entails practically no assumptions, save
for the existence of conditional distributions and the thesis that
conditional expectations are ``reasonable'' or ``meaningful'' as predictors
of unknowns. A \textit{dynamic statistical model} builds upon the theme of equation
(5.3) by parameterizing the $V(t)$ process. One such parameterization is to
let $V(t)=\alpha y(t-1)$, for some constant $\alpha>0$. This parameterization
states that the expected change in $y(t)$, namely $y(t)-y(t-1)$ is
proportional to $y(t-1)$, with $\alpha>0$ as the constant of proportionality. 
With this in place it is easy to see that
\begin{equation*}
y(t)-y(t-1)=\alpha y(t-1)+\bigl(M(t)-M(t-1)\bigr)\text{,}
\end{equation*}
or that
\begin{equation*}
y(t)=(1+\alpha)\,y(t-1)+\bigl(M(t)-M(t-1)\bigr)\text{,}\tag*{(5.6)}
\end{equation*}
an autoregressive process of order 1, with orthogonal errors having mean zero
(the latter property is known as \textit{colored noise}). Note that $y(t)$ is a stationary process only when $-2<\alpha<0$.

Since a simplified version of Kalman's state space model of equation (4.1)
is of the form
\begin{equation*}
y(t)=s(t)+n(t)\text{, and}
\end{equation*}
\begin{equation*}
s(t)=s(t-1)+a(t)\text{,}\tag*{(5.7)}
\end{equation*}
a correspondence between the above and the model based on Doob's
decomposition$-$equation (5.3)$-$is easy to identify. Specifically,
iterating on equation (5.7), it is easy to see that for any $n\leq t$, 
\begin{equation*}
y(t)=s(t-n)+n(t)+\sum_{j=t-n+1}^ta(j)\text{,}
\end{equation*}
so that for $n=t$,
\begin{equation*}
y(t)=\bigl(s(0)+n(t)\bigr)+\sum_{j=1}^ta(j).
\end{equation*}
The desired correspondence holds if $\bigl(s(0)+n(t)\bigr)$ is identified with
$M(t)$, and $a(j)$ identified with $V(j)$. It is assumed that at $t=0$, the
value of the signal $s(0)$ is known.

There exists a continuous version of the Doob composition, known as the
\textit{Doob-Meyer Decomposition}, which spawns a martingale process $\{M(t);\,t\geq
0\}$ with respect to the process $\{y(t);\,t\geq 0\}$. A consequence of the
martingale process is an ability to use Levy's Theorem $[$cf. Doob (1953), Theorem 11.9$]$, which
asserts that a martingale process with variance $t$ is a Brownian motion
process (also known as Wiener process). 
Results such as these, expand the
scope of Wiener's theory by enabling filtering under more general Gaussian processes, 
non-Gaussian, and discontinuous processes. For example, Kara, Mandrekar, and Park (1974) discuss recursive least-squares estimation when the noise is a martingale, and Mandrekar and Naik-Nimbalkar (2009) consider estimation when the noise is a fractional Brownian  motion. The recent books by Mandrekar and Rudinger (2015), and Mandrekar and Gawarecki (2015) outline the theory and provide a source of references.

Whereas all of the above is conceptually natural, implementation poses a challenge. As a consequence, the filtering algorithm which bears Kalman's name, continues to be used and discussed.

\subsection{Antecedents to Kalman's Filtering Algorithm}

The algorithm proposed by Kalman (1960), even for a simplified linear version of the
state-space model, is cumbersome to describe. The essential features of this algorithm are: all data available up to some time are employed to estimate the state parameter at that time; at any given time one does not retain the whole record of all observations up to that time, their effect being encapsulated in the estimate of the state vector at that time; new data are optimally combined with the most recent state vector.  In Section 6 we overview a
Bayesian prior to posterior iterative approach for addressing the filtering,
smoothing, and prediction as prescribed by equation (5.7). Many find the
Bayesian perspective easier to digest. But before doing so we outline below
some antecedents that may have lead Kalman to develop his algorithm $[$cf.
Sorenson (1970)$]$.

For the set-up of (5.7), a filter's weighting function for signal
$s(t)$, can be easily developed via the method of least squares based on $n$
previous observations $y(t)$. However, a new solution needs to be generated
for each new observation and this could be demanding. The idea that upon the
receipt of $y(n+1)$, an estimate of $s(n+1)$ can be based on an estimate of
$s(n)$ obtained via $y(1),\ldots,y(n)$, is due to Folin in 1955 \lbrack cf. Bucy
(1968), p. 129\rbrack. The notion of recursive filtering and prediction is
also present in the works of Swerling (Jan. 1958) and Blum (March 1958), though Swerling's set up is deterministic and there is no mention by him of state space models. Swerling's work was motivated by applications to estimating orbits of earth satellites and space vehicles. A comparison of Swerling's and Kalman's formulation is in Swerling (1998).  In
the statistical sciences the method of stochastic approximation by Robbins
(1951) and by Kiefer and Wolfowitz (1952), were also being studied. Thus it
appears that the time was ripe for the recursive approach to state-space
estimation proposed by Kalman in 1960$-$albeit almost after 9 years since the
works of Robbins and Kiefer and Wolfowitz. According to Sorenson (1970),
Swerling in 1968 wrote a letter to the AIAA Journal claiming priority for the
Kalman filter equations based on his 1958 work described in a RAND memorandum
on orbit determination.

Of noteworthy mention here is also the striking work (in the former USSR) of Stratonovich (1959, 1960a, 1960b). Stratonovich was the first to emphasize the importance of Markov processes in signal detection in continuous time, and in the sequel, the development of the theory of Conditional Markov Processes.

\section{Bayesian Learning and (Kalman) Filtering}

Bayesian learning via a prior to posterior iteration can be seen as an
implementation of the conditional expectation principle, which is the basis of
Doob's decomposition. When the underlying distributions are assumed to be Gaussian  
(or more generally spherically symmetric) and admit a state-space representation, the principle of least squares and
conditional expectation yield identical answers. To appreciate this and
related matters, we find it convenient to re-cast the state-space model of
equation (5.7) in a notation palatable to statisticians \lbrack eg. Meinhold
and Singpurwalla (1983)\rbrack\ as:
\begin{equation*}
Y_t=\theta_t+v_t\tag*{(6.1.a)}
\end{equation*}
\begin{equation*}
\theta_t=\theta_{t-1}+w_t\text{,}\tag*{(6.1.b)}
\end{equation*}
where $\theta_t$ is an unknown (dynamic) parameter whose value changes with
$t=0,1,2,\ldots,$ and $v_t$ and $w_t$ are errors. The $v_t$'s are assumed to
be uncorrelated and identically normally distributed with mean $0$, and
variance $\sigma_v^2$; this is denoted as $v_t\sim\mathcal{N}(0,\sigma_v^2)$,
with $v_t$ independent of $w_t$, and $w_t\sim\mathcal{N}(0,\sigma_w^2)$. If
$\mathbf{Y_t}=(Y_1,Y_2,\ldots,Y_t)$, then the prediction
problem boils down to assessment of
$P(Y_t|\mathbf{Y_{t-1}})$, the conditional distribution of a
future $Y_t$, \underline{were} (supposing that) $\mathbf{Y_{t-1}}$ be known. Note the
emphasis on the word ``were.'' By contrast Wiener's prediction problem boils
down to an assessment of $P(Y_t|\mathbf{y_{t-1}})$, the
distribution of a future $Y_t$ having \underline{actually} observed
$\mathbf{y_{t-1}}=(y_1,\ldots,y_{t-1})$, where $y_\tau$ is an
observed realization of $Y_\tau$; note the emphasis on the word "actually." The distinction between
$P(Y_t|\mathbf{Y_{t-1}})$ and
$P(Y_t|\mathbf{y_{t-1}})$ is philosophical and subtle. The
mechanics leading to an assessment of both \underline{could} be the same, but this need not be so; see Section
6.2. Similarly with filtering and smoothing, which entail assessments of
$P(\theta_t|\mathbf{Y_t})$ and
$P(\theta_j|\mathbf{Y_t})$, respectively, for any
$j=1,2,\ldots,(t-1)$. What follows next is merely an application of the
calculus of probability to achieve the desired assessments. Specifically:
\begin{equation*}
P(Y_t|\mathbf{Y_{t-1}})=\int_{\theta_t}P(Y_t|\theta_t,\mathbf{Y_{t-1}})\,%
P(\theta_t|\mathbf{Y_{t-1}})\,d\theta_t=\int_{\theta_t}P(Y_t|\theta_t)\,P(\theta_t|\mathbf{Y_{t-1}})\,d\theta_t, \tag*{(6.2)}
\end{equation*}
by law of total probability, and assuming $Y_t$ independent of $\mathbf{Y_{t-1}}$. 

From (6.1.a), $(Y_t|\theta_t)\sim\mathcal{N}(\theta_t,\sigma_v^2)$, so that to complete
the assessment of $P(Y_t|\mathbf{Y_{t-1}})$ we need to know
$P(\theta_t|\mathbf{Y_{t-1}})$. By extending the
conversation to $\theta_{t-1}$, and then assuming, given $\theta_{t-1}$,  $\theta_t$ independent of
$ \mathbf{Y_{t-1}}$ 
\begin{equation*}
P(\theta_t|\mathbf{Y_{t-1}})=\int_{\theta_{t-1}}P(\theta_t|%
\theta_{t-1})\,P(\theta_{t-1}|\mathbf{Y_{t-1}})\,d%
\theta_{t-1}.\tag*{(6.3)}
\end{equation*}

To obtain $P(\theta_t|\theta_{t-1})$, we lean on (6.1.b) to assert
that $(\theta_t|\theta_{t-1})\sim\mathcal{N}(\theta_{t-1},\sigma_w^2)$. With
the above in place, suppose that 
$P(\theta_{t-1}|\mathbf{Y_{t-1}})$ is governed by 
$(\theta_{t-1}|\mathbf{Y_{t-1}})\sim\mathcal{N}(m_{t-1}$,$\,%
C_{t-1})$, then by the properties of the Gaussian distribution,
$(\theta_t|\mathbf{Y_{t-1}})\sim\mathcal{N}(m_{t-1}$,$\,%
C_{t-1}+\sigma_w^2)$. Using an analogous argument $P(Y_t|\mathbf{Y_{t-1}})$ 
of equation (6.2) is governed by 
$\mathcal{N}(m_{t-1}$, $\,C_{t-1}+%
\sigma_w^2+\sigma_v^2)$. 

Were we to receive the next observation $Y_t$, then we would be required to
assess $P(Y_{t+1}|\mathbf{Y_t})$, and to do so we would need
to know $P(\theta_t|\mathbf{Y_t})$. By Bayes' Law, 
\begin{equation*}
P(\theta_t|\mathbf{Y_t})=P(\theta_t|Y_t, 
\mathbf{Y_{t-1}})\propto\mathcal{L}(\theta_t;Y_t)P(\theta_t|\mathbf{Y_{t-1}}),\tag*{(6.4)}
\end{equation*}
where $\mathcal{L}(\theta_t;Y_t)$ is the likelihood of $\theta_t$ with $Y_t$
fixed (and assumed to depend on $Y_t$ alone). The last term  is 
$(\theta_t|\mathbf{Y_{t-1}})\sim\mathcal{N}(m_{t-1}$,$\,%
C_{t-1}+\sigma_w^2)$. Assuming that the likelihood
$\mathcal{L}(\theta_t;Y_t)$ is induced by the feature that
$(Y_t|\theta_t)\sim\mathcal{N}(\theta_t,\sigma_v^2)-$equation (6.1.a)$-$it
follows from routine Bayesian prior to posterior calculations that
$(\theta_t|\mathbf{Y_t})\sim\mathcal{N}(m_t$,$\,C_t)$.

Wiener's problem of \textit{smoothing} boils down to an assessment of $\theta_t$, were
we to know $ \mathbf{Y_{t+1}}$. For this we assess
$P(\theta_t,\theta_{t+1}| \mathbf{Y_{t+1}})$ and 
integrate out $\theta_{t+1}$. However,
$P(\theta_t,\theta_{t+1}|\mathbf{Y_{t+1}})$ can be
obtained as a conditional distribution of
$P(\theta_t,\theta_{t+1},Y_{t+1}| \mathbf{Y_t})$, and
this can be assessed via 
$P(\theta_t|\mathbf{Y_t})$, $P(Y_{t+1}|\mathbf{Y_t})$, and
$P(\theta_{t+1}|\mathbf{Y_t})$, all of which we are
able to obtain via the discussions of the previous paragraphs. Smoothing pertains to making revised probabilistic assessments of $\theta_t$ given all the currently observed information. The intuition here is that better estimates of $\theta_t$ are obtained when data subsequent to $Y_t$ is also at hand.

Thus under the simple set-up of equation (6.1), Wiener's filtering, smoothing
and prediction problems can be solved in closed form via the principle of
conditional expectation, implemented via the mechanics of Bayesian learning.
The process of predict and update provides an optimal Bayesian solution for the linear Gaussian state-space model. Matters become computationally challenging when the error distributions are non-Gaussian, have non-constant variances, are correlated, or when (6.1) entails
non-linearities. When such is the case, one
resorts to \textit{Gibbs sampling} which is a \textit{Markov Chain Monte Carlo} (MCMC)
method; it is outlined in Section 6.1. The efficacy of MCMC depends on the convergence of a Markov Chain to
an equilibrium distribution. The essence of the Gibbs sampling as applied to a linear Gaussian state-space model (the Kalman filter model) is described  below. Our aim is to
set the stage for a discussion of the particle filtering algorithm as an
alternative to the Gibbs sampling. But before doing so, it may be helpful to remark that if the observed process is "invertible" in the sense of Box and Jenkins (1970), then a naive approach for overcoming the obstacle of a growing dimension is to \textit{filter out} (i.e. eliminate) observations that have occurred in the remote past. When the process is not invertible, then the naive approach will lead to misleading answers. An archetypal example of a non-invertible process is a moving average process of order one, whose coefficient is greater than or equal to one in absolute value. Non-invertibility can arise due to over differencing; see for example, Abraham and Ledolter (1983, pp. 233- 236).

Prior to the advent of Gibbs sampling, the matter of nonlinearity (i.e. an inability to write the evolution of the state variable and/or the observed process as a linear model) was treated by variants of the procedure described above, via what is known as an \textit{extended Kalman Filter} (EKF). This entailed a local linearization of the nonlinear equations by a Taylor series approximation [cf. Anderson and Moore (1979)]. Since Swerling's (1959) original formulation included   
the nonlinear case as well, it may be claimed that the EKF is the original Swerling filter. However, the EKF was found to be credible only under scenarios wherein the underlying nonlinearities were almost linear, and thus an alternative, namely, the \textit{unscented Kalman filter} (UKF) was proposed by Julier and Uhlmann (1977); also see van der Merwe et al. (2000). The UKF which is not restricted to the requirement of Gaussian distributions is based on the intuition that it is easier to approximate a Gaussian distribution than it is to approximate an arbitrary nonlinear function. Accordingly, a set of points that are deterministically selected from the Gaussian approximation to $P(\theta_t|\mathbf{Y_t})$ are propagated through the underlying nonlinearity, and the points thus propagated used to obtain a Gaussian approximation to the transformed distribution [cf. Arulampalam, Maskell, Gordon and Clapp (2002)]. If the underlying density is bimodal or heavily skewed, then a Gaussian will not approximate it well spawning the need for \textit{robustifying} the Kalman filter using influence functions or thick tailed distributions, such as the Student's$-$t [cf. Meinhold and Singpurwalla (1989)], or by Monte Carlo based approaches such as Gibbs sampling or particle filtering.     

\subsection{The Gibbs Sampling Algorithm for Kalman Filtering}
As mentioned in the previous section, the Gibbs sampling algorithm for Kalman filtering becomes germane under non-Gaussianity and non-linearity of the Kalman filter model for which a closed form solution exists when otherwise.
A fundamental step in the Kalman filter algorithm is the recursive
transitioning from $P(\theta_{t-1}|\mathbf{Y_{t-1}})$ to
$P(\theta_t|\mathbf{Y_t})$. This operation entails a
likelihood and an application of Bayes' law. The specifics of the operation
were outlined in the paragraph following (6.4). There is an important
aspect of this operation, which is germane to particle filtering. Specifically, to transition from
$P(\theta_{t-1}|\mathbf{Y_{t-1}})$ to
$P(\theta_t|\mathbf{Y_t})$, one first propagates from $\theta_{t-1}$, to
$\theta_t$ via (6.1.b), and then brings in the
effect of $Y_t$ via the likelihood and Bayes' Law. An exercise like this is
legislated by a factorization of the form
\begin{equation*}
P(Y_{t+1}\text{,}\,\theta_{t+1}|\theta_t)=P(Y_{t+1}|\theta_{t+1})P(%
\theta_{t+1}|\theta_t)\text{,}
\end{equation*}
assuming that given $\theta_{t+1}$, $Y_{t+1}$ is independent of $\theta_t$.
Thus with conventional filtering, the motto is: ``\textit{propagate
first$-$update next,}'' a meaningful thing to do when a real
time decision is to be made at $t$, based on knowledge about $\theta_t$
at time $t$; for example, in automatic control. Here all that matters is
$P(\theta_t|\mathbf{Y_t})$. 

However, were the 
scenario be such that a decision based on knowledge about $\theta_t$ can be
delayed until time $(t+1)$ with $\mathbf{Y_{t+1}}$ at hand,
then a smoothed assessment of $\theta_t$ based on
$\mathbf{Y_{t+1}}$ would be preferable to one based on
$\mathbf{Y_{t}}$ alone. Such delayed decision scenarios arise
in the context of statistical inference. Such an exercise is legislated by the factorization 
\begin{equation*}
P(Y_{t+1}\text{,}\,\theta_{t+1}|\theta_t)=P(\theta_{t+1}|\theta_t\text{,}\,%
Y_{t+1})P(Y_{t+1}|\theta_t),
\end{equation*}
where the motto would be to: ``\textit{update first$-$propagate
next.}'' This motto does
not entail any assumption of conditional independence. Either motto can be
implemented in both the Gibbs sampling algorithm, or the particle filter
mechanism (discussed in Section 7). The expression
$P(Y_{t+1},\theta_{t+1}|\theta_t)$ which arises in the context of
transitioning from $P(\theta_t|\mathbf{Y_{t}})$ to
$P(\theta_{t+1}|\mathbf{Y_{t+1}})$ will be motivated in
Section 7.1.

As a synopsis of the Gibbs sampling algorithm for the model of
(6.1), we focus attention on the case $t=2$, and suppose that $Y_1$ and
$Y_2$ are observed as $y_1$ and $y_2$, respectively. Consider the 4-tuple
$(\theta_1$,$\,\theta_2$,$\,y_1$,$\,y_2)$. The set of 2 conditional
distributions spawned by this 4-tuple have the distributions: 
\begin{center}
$(\theta_1|\theta_2$,$\,y_1$,$\,y_2)\sim(\theta_1|\theta_2$,$\,y_1)$

$(\theta_2|\theta_1$,$\,y_1$,$\,y_2)\sim(\theta_2|\theta_1$,$\,y_2)$.
\end{center}
\vspace{.1in}

Setting $\theta^{(0)}_1$ and $\theta^{(0)}_2$ as starting values of $\theta_1$ and
$\theta_2$, we update $\theta^{(0)}_2$ to $\theta^{(1)}_2$ by generating a sample (indeed, a particle)
from $(\theta_2|\theta^{(0)}_1$,$\,y_2)$. To do so, we note that
\begin{equation*}
P(\theta_2|\theta^{(0)}_1\text{,}\,y_2)\propto
P(y_2|\theta_2,\theta^{(0)}_1)\,P(\theta_2|\theta^{(0)}_1)=P(y_2|\theta_2)\,P(%
\theta_2|\theta^{(0)}_1)\text{,}
\end{equation*}
and the last two probabilities are specified by the assumed structure of
(6.1). 

Next, we generate $\theta^{(1)}_1$ from
$P(\theta_1|\theta^{(1)}_2$,$\,y_1)\propto$
$P(\theta^{(1)}_2|\theta_1$,$\,y_1)\,\mathcal{L}(\theta_1;\,y_1)\,P(\theta_1)$ $=P(%
\theta^{(1)}_2|\theta_1)\mathcal{L}(\theta_1;\,y_1)\,P(\theta_1)$, where
$\mathcal{L}$ is the likelihood. The $\theta^{(1)}_1$ is an
update of $\theta^{(0)}_1$.

The above process repeats, so that after $k$ iterations we have 
\begin{equation*}
(\theta^{(0)}_1,\text{ }\theta^{(0)}_2),\,(\theta^{(1)}_1,\text{
}\theta^{(1)}_2),\,(\theta^{(2)}_1,\text{ }\theta^{(2)}_2),\ldots,\,(\theta^{(k)}_1,\text{
}\theta^{(k)}_2)
\end{equation*}
based on the starting values $\theta^{(0)}_1$ and $\theta^{(0)}_2$, and given values
$y_1$ and $y_2$. Under some mild regularity conditions, as
$k\rightarrow\infty$, the distribution of $(\theta^{(k)}_1,$ $\theta^{(k)}_2)$
converges to the posterior distribution $P(\theta_1,\theta_2|\,y_1$,$\,y_2)$;
see Gelfand and Smith (1990). Alternatively, samples from the posterior
distribution $P(\theta_1,\theta_2|\,y_1$,$\,y_2)$ can also be generated using the
\textit{forward filtering backward sampling algorithm}; see Fruhwirth-Schnatter
(1994), and Carter and Kohn (1994).

Cleary, each new observation increases the size of
the tuple by two, and calls for the generation of a new set of $k$ variates.
This is computationally burdensome which the particle
filter avoids. But first some words about the caveat of conditioning.

\subsection{Filtering, Smoothing, and the Principle of Conditionalization}

An important, but underemphasized, point pertains to be the subjunctive nature of the
discussion up until now. This has to do with the feature that all of probability, to include conditional probability and Bayes' Law, is in the subjunctive mood. That is, the discussion up until now is based on the premise that ``\underline{were}  $Y_i$
to be observed as $y_i$, $\,i=1,2,\ldots,t$,'' and \underline{not} on the premise that $Y_i$ is \underline{actually} observed as $y_i$.  The above difference is encapsulated in the claim that all of probability is in the irrealis (or subjunctive) mood, whereas with actual data at hand, inference has to be in the indicative mood; see, for example, Singpurwalla (2016). The development of Section 6.1 and the ensuing histograms therein are meaningful for some assumed value $y_i$ of $Y_i$. What happens to this development if when the $y_i$'s are the actual
observed values of $Y_i$'s ?

Our answer is that everything that has been said before 
continues to be valid, but only if the philosophical 
\textit{principle of conditionalization} is adopted \lbrack cf. Diaconis and Zabell (1982), or
Singpurwalla (2007)\rbrack. This means that underlying the current practice
in signal processing, forecasting, and control theory, there is an implicit adherence to 
the principle of conditionalization. Making this point explicit to the engineering and the statistical communities is a feature of this paper which goes beyond a mere review. The principle of conditionalization is best exposited via a subjectivistic interpretation of probability. 

Suppose that for two uncertain events $A$ and $B$, one is able to specify the conditional probability $P(A|B)$. In the subjectivistic context $P(A|B)$ denotes a two-sided bet on the occurrence of event $A$, under the supposition that event $B$ has occurred. Under the principle of conditionalization, the above bet must continue to hold even when one is informed that event $B$ has actually occurred. In other words, under  conditionalization, one's disposition towards betting on event $A$ is indifferent as to whether $B$ is supposed to have occurred or has actually occurred. Several individuals starting with Ramsey (1931) have questioned the universality of this principle. They have claimed that the actual occurrence of $B$ could change one's disposition to bets on event $A$ made under the supposition that event $B$ has occurred. In statistical inference, using Bayes' Law, the principle of conditionalization manifests itself whenever the likelihood is specified by interchanging the roles of parameters and the variables in an assumed probability model. This practice is so routinely followed that its philosophical underpinnings are almost forgotten.

Were the principle of conditionalization not adopted, then the likelihood
of (6.4) would not necessarily be induced by the feature that
$(Y_t|\theta_t)\sim\mathcal{N}(\theta_t,\sigma_v^2)$, and the commonly used
Kalman filter equations for filtering, smoothing, and extrapolations would
not follow ! Under such circumstances, the likelihood$-$which is a weighting
function$-$will be an arbitrary function, specified via judgmental
considerations, and the ensuing relationships different from those used in
current practice.

\section{The Particle Filter}

As stated, a closed form solution to the Kalman Filter Model assuming
that the underlying distributions are Gaussian entails the inversion of
matrices whose size grows with the number of observations. This, in the
current era of big data can be forbidding. Gibbs sampling can come to the
rescue here, but the Gibbs sampler can be computationally burdensome and its success rests on the convergence of
the underlying Markov Chain to an equilibrium distribution \lbrack cf. Smith
and Roberts (1993)\rbrack. By contrast the particle filter mechanism mimics
the Bayesian prior to posterior learning step by step, and leans on the law
of large numbers to ensure convergence. Indeed, the particle filter (also known as a \textit{genetic Monte Carlo} algorithm) better
exploits the Markovian nature of equation (6.1.a) than the Gibbs sampling
algorithm, and in so doing it: 

i) Circumvents the computational burden spawned by the growing size of the
MCMC tuple$-$see section 6.1, and

ii) Obviates the need for large storage memory by not requiring that all observations prior to the current $y_t$ be retained. This feature of
particle filtering truly embodies the essential spirit of recursive
estimation as enunciated by Folin in 1955. But as will be pointed out later, the particle filter is not without its drawbacks.

Particle filters (PF) work online and use a discrete set of values called \textit{particles}, each with a weight, to represent the distribution of a state at time $t$, and to update this distribution at each subsequent time by changing the particle weights according to their likelihoods. There are several versions of the PF, and several surveys and tutorials about it, one of the most comprehensive one being that by Arulampalam, Maskell, Gordon, and Clapp (2002), and one of the most recent one being that by Doucet and Johansen (2011). Also noteworthy is the exhaustive treatise by Chen (2003), and the expository set of lecture notes by Pollock (2010) and by Turner (2013). Chen's (2003) paper is all inclusive with a very thorough set of references; it is written from the perspective of a control theorist with an emphasis on engineering mathematics which statisticians may find challenging to decipher. The current paper may serve as a good prelude to Chen's paper for those who are interested in digesting the material therein.

 \textit{Importance sampling} (IS) and its variants are the key tools which drive the PF; thus it seems appropriate to give below a broad based overview of the essentials of IS.

\subsection{Importance Sampling and its Variants}

IS was originally introduced in statistical physics as a variance reducing technique. The essence of the idea here is that a mathematical expectation with respect to a target distribution is well approximated by a weighted average of random draws from another distribution called the \textit{importance distribution}. That is, if a random variable $\theta$ has a probability density $p(\theta)$, then 
$$
\mu_{f}=E_{p}[f(\theta)]=\int f(\theta) p(\theta) d\theta,
$$
and if $q(\theta)$ is some other probability density of $\theta$, with the property that $q(\theta)>0$, whenever $f(\theta) p(\theta)\neq0$, then $\mu_{f}=E_{q}[\omega(\theta)f(\theta)]$, where $\omega(\cdot)=p(\cdot)/q(\cdot)$. The presumption here is that it is possible to sample from $q(\theta)$ but not from $p(\theta)$.

Thus, if we draw a sample $\theta^{(1)},\ldots,\theta^{(m)}$ from $q(\theta)$, then $\sum_{i}^{m} f(\theta^{(i)}) \omega(\theta^{(i)}) $ will (by the strong law of large numbers) converge almost surely to $\mu_{f}$.  

The merit of IS is clearly apparent in a Bayesian context wherein the posterior, $p(\theta|y)\propto \mathcal{L}(\theta;y)q(\theta)$, is known only up to a normalizing constant, so that it is possible to sample from the prior $q(\theta)$ but not from $p(\theta|y)$. When such is the case, an estimate of $\mu_{f}=\int_{\theta} f(\theta;y)p(\theta|y) d\theta$ is given by 
$$
\hat{\mu}_{f}=\sum_{i} f(\theta^{(i)};y)\omega(\theta^{(i)}),
$$
where $\omega(\theta^{(i)})=\frac{\mathcal{L}(\theta^{(i)};y)}{\sum_{j} \mathcal{L}(\theta^{(j)};y)} $ and $\theta^{(1)},\theta^{(2)},\ldots $ is a random draw from $q(\theta)$.

In state-space models $\theta$ is high dimensional and $p(\theta)$ leads to a chain like decomposition of $\theta$. This enables the sequential construction of the importance density, and now one is able to sequentially update the posterior density at some time $t$, without modifying the previously simulated states. This is the idea behind \textit{sequential importance sampling} (SIS) discussed by Liu (2001). A common problem encountered with SIS is the \textit{degeneracy} phenomenon, where after few iterations all but a few particles will have negligible weights. Indeed it is shown by Liu (2001) that the weight sequence forms a martingale leading to the feature that the variance of the importance weights increase over time. Consequently, a very small portion of the draws carry most of the weight making the SIS procedure computationally inefficient. Details about the above matters are given in Kong, Liu, and Wong (1994), who among other things propose an approach for overcoming the problem of degeneracy.

Resampling is another approach by which the effects of degeneracy can be reduced. The idea here is to eliminate particles having a small weight and concentrate on particles with a large weight by picking a particle with a probability proportional to its weight. Such a particle filtering process was proposed by Gordon, Salmond, and Smith (1993) in their classic and ground breaking paper; it is known as \textit{sampling importance resampling} (SIR). Whereas the SIR filter resamples particles at the end of an iteration, say at time $(t-1)$ before an observation $y_t$ at $t$ is taken, the \textit{auxiliary particle filter} (APF) introduced by Pitt and Shephard (1999), employs the knowledge about $y_t$ before resampling at $(t-1)$. This ensures that particles that are likely to be compatible with $y_t$ have a good chance of surviving, and in so doing makes the particle filtering process computationally efficient. 

Collectively, the process of using a discrete set of weighted particles to represent the distribution  of a state, and to update this distribution by changing the particle weights, as is done under the SIS, SIR, and APF algorithms is also known as \textit{sequential Monte Carlo} (SMC), a term coined by Liu and Chen (1998). The PF methods mentioned above suffer from the "curse of dimensionality" [cf. Bengtsson, Bickel, and Li (2008)]. This happens when $p-$ the dimension of the state space, and $q-$ the dimension of the observation vector are very large in relation to $n$, where $t=1,2,\ldots,n$. When such is the case, which arises in the context of climate modeling, dimension reduction techniques which entail a decomposition of the state and observation vectors into many overlapping patches, are invoked. The \textit{ensemble Kalman filter} (EnKF), which is a combination of SMC and the Kalman filter, works under the decomposition scheme mentioned above, whereas the PF does not [cf. Lei and Bickel (2009)].

\subsection{Architecture(s) of the Particle Filter Algorithm}  

Figures 4 and 5 encapsulate the architecture of two versions of 
the particle filtering
mechanism, the former subscribing to the motto of propagate first$-$update
next, and the latter to that of update first$-$propagate next; see Section 6.1. These figures
can be construed as a graphical appreciation of the particle filter
mechanism. The essential import of the mechanism of Figures 4 and 5 pertains to
the process of transitioning from the distribution of
$(\theta_t|\mathbf{Y_t})$ to the distribution of
$(\theta_{t+1}|\mathbf{Y_{t+1}})$ upon the receipt of new
data. For \underline{convenience} and ease of exposition, we assume that
$(\theta_t|\mathbf{Y_t})\sim\mathcal{N}(m_t$,$\,C_t)$; the
notation used here is that of Section 6. The mechanics of the particle filter algorithm
is general enough to accommodate distributions other than the Gaussian, and
that is another virtue.

As a matter of historical note, even though the recent impetus in particle
filtering has been triggered by the 1993 paper of Gordon, Salmond and Smith,
the core of the underlying idea goes back to Galton (1877) \lbrack cf.
Stigler (2011)\rbrack.

To discuss the transitioning from a specified
$P(\theta_t|\mathbf{Y_t})$ to a
$P(\theta_{t+1}|\mathbf{Y_{t+1}})$ upon receipt of
$Y_{t+1}$, we start by considering $P(\theta_{t+1}|\mathbf{Y_{t+1}})=P(\theta_{t+1}|\mathbf{Y_t},Y_{t+1})\propto
P(\theta_{t+1},Y_{t+1}|\mathbf{Y_t})$, and observe that
\begin{equation*}
P(\theta_{t+1},Y_{t+1}|\mathbf{Y_t})=\int_{%
\theta_t}P(\theta_{t+1},Y_{t+1}|\theta_t,\mathbf{Y_t})P(\theta_t|\mathbf{Y_t})d\theta_t.
\end{equation*}

Since $P(\theta_t|\mathbf{Y_t})$ is assumed known as
$\mathcal{N}(m_t$,$\,C_t)$, we focus attention on $P(\theta_{t+1},Y_{t+1}|\theta_t,\mathbf{Y_t})$ to see how it could be simplified by factorization. There are two factorizations of this joint conditional distribution each leading to a protocol for updating.  The first
factorization leads to the protocol of \textit{propagate first - update next}; the second to the \textit{update first - propagate next} protocol; see Section 6.1.

\subsubsection{The Propagate First$\boldsymbol{-}$Update Next Protocol }

The entity
$P(\theta_{t+1},Y_{t+1}|\theta_t,\mathbf{Y_t})$ of the equation above can
be factored as
$P(Y_{t+1}|\theta_{t+1},\theta_t,\mathbf{Y_t})\times
P(\theta_{t+1}|\theta_t,\mathbf{Y_t})$.
If $Y_{t+1}$ is assumed independent of $\theta_t$ and
$\mathbf{Y_t}$ given $\theta_{t+1}$, and $\theta_{t+1}$ assumed independent of $\mathbf{Y_t}$ given $\theta_{t}$,
then this factorization simplifies as:
\begin{equation*}
P(\theta_{t+1},Y_{t+1}|\theta_t,\mathbf{Y_t})=P(Y_{t+1}|\theta_{t+1})P(\theta_{t+1}|\theta_t).\tag*{(7.1)}
\end{equation*}
Equation (7.1) is the basis of the "propagate first-update next"
protocol. By this it is meant that in moving from the right to left of this equation, one starts by propagating
$\theta_t$ to $\theta_{t+1}$ via equation (6.1.b), and then upon the receipt of
$Y_{t+1}$ updates $\theta_t$ to $\theta_{t+1}$ [using the expression (7.2) given below].

Plugging the simplified factorization of equation (7.1) in the expression for $P(\theta_{t+1}|\mathbf{Y_{t+1}})$ discussed before, we have
\begin{equation*}
P(\theta_{t+1}|\mathbf{Y_{t+1}})\propto   \int_{%
\theta_t}P(Y_{t+1}|\theta_{t+1})P(\theta_{t+1}|\theta_t)P(\theta_t|\mathbf{Y_t})d\theta_t. \tag*{(7.2)}
\end{equation*}
The essence of particle filtering under this propagate first-update next protocol is 
an implementation of equation (7.2), via
a simulation exercise, wherein one starts by generating a sample of size $N$
from the distribution of $(\theta_t|\mathbf{Y_t})$, which for purposes of discussion has been assumed Gaussian,  and works one's way from right to left. Denote these generated values, known as \textit{particles}, by $\theta^{(i)}_t$,$\,i=1,\ldots,N$; these particles get propagated to $\theta^{(i)}_{t+1}$ via the mechanism driving $P(\theta_{t+1}|\theta_t)$, namely, equation (6.1). With $Y_{t+1}$ observed as $y_{t+1}$, $P(Y_{t+1}|\theta_{t+1})$ gets replaced by $\mathcal{L}(\theta_{t+1};y_{t+1})=P(y_{t+1}|\theta_{t+1})$, the "filtering" likelihood of $\theta_{t+1}$ under an observed $y_{t+1}$. The rest follows from the schematics of Figure 4, with equation (7.2) in perspective. The \textit{importance weights} $w^{(i)}_{t+1}=\frac{P(y_{t+1}|\theta^{(i)}_{t+1})}{\sum_{j=1}^NP(y_{t+1}|\theta^{(j)}_{t+1})}$ modulate the propagated particles $\theta^{(i)}_{t+1}$ by
emphasizing those  which are meaningful, and diffusing
those which are skewed (i.e. outliers); these importance weights add to 1.  Despite the introduction of these
modulating weights, there remains the possibility of degeneracy, because the
generated particles could be concentrated around a few values causing a
collapse of the process. As mentioned, this is a drawback of all such simulation exercises. Recall, the
importance weights sum to one, and each weight is proportional to the
likelihood of the $\theta^{(i)}_{t+1}$ which spawns it \lbrack cf. Carvalho
et al. (2010)\rbrack. 
\begin{figure}[!ht]
\begin{center}
\includegraphics[scale=0.75]{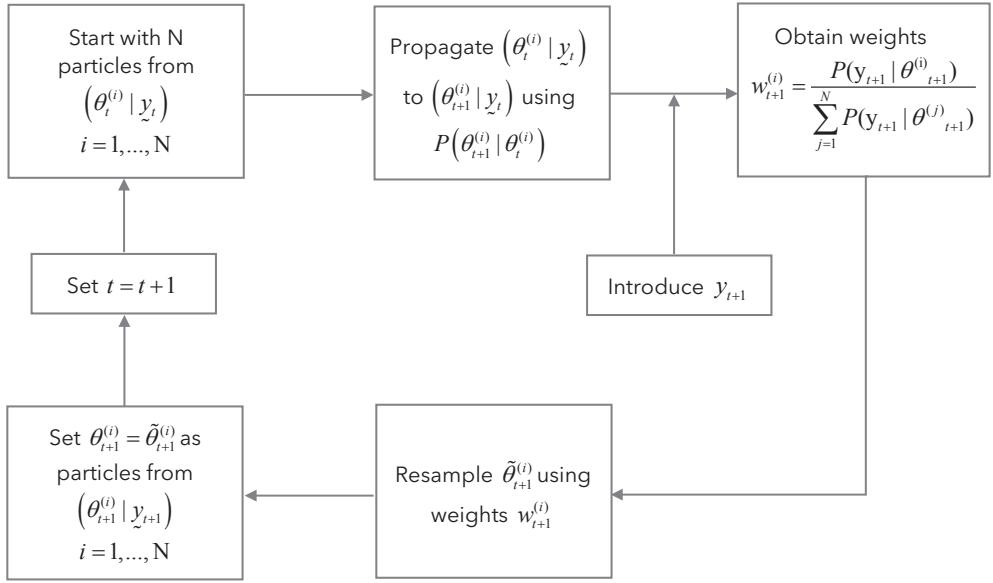}
\end{center}
\label{f4}
\caption{Particle Filtering under propagate First$-$Update Next Protocol.}
\end{figure}
Observe that the flow of actions depicted in Figure 4 mimics the
architecture of equation (7.2) as one moves from its right to its left.

The one open question pertains to $N$ the number of cycles that the
algorithm needs to execute. Barring the prospect of degeneracy, the law of
large numbers will, for large $N$, ensure convergence to a stationary
distribution. This stationary distribution represents the updated (posterior)
distribution $\mathcal{N}(m_{t+1}$,$\,C_{t+1})$ in our assumed case.

\subsubsection{The Update First$\boldsymbol{-}$Propagate Next Protocol }

The entity
$P(\theta_{t+1},Y_{t+1}|\theta_t,\mathbf{Y_t})$ of
the previous section has an alternate factorization, and this factorization forms
the basis of the "update first$-$propagate next protocol" for the particle filter. Thus, the two
protocols of particle filtering discussed here are motivated by the two
factorizations of $P(\theta_{t+1},Y_{t+1}|\theta_t,\mathbf{Y_t})$. Specifically, $P(\theta_{t+1},Y_{t+1}|\theta_t,\mathbf{Y_t})$, can also be factored as follows:
\begin{equation*}
P(\theta_{t+1},Y_{t+1}|\theta_t,\mathbf{Y_t})=P(\theta_{t+1}|\theta_t,Y_{t+1},\mathbf{Y_t})P(Y_{t+1}|\theta_t,\mathbf{Y_t})=
P(\theta_{t+1}|\theta_t,Y_{t+1})P(Y_{t+1}|\theta_t,\mathbf{Y_t})\tag*{(7.3)},
\end{equation*}
if $\theta_{t+1}$ is assumed independent of $\mathbf{Y_t}$,
given $\theta_t$ and $Y_{t+1}$.

Incorporating the factorization of (7.3) into the relationship
\begin{equation*}
P(\theta_{t+1}|\mathbf{Y_{t+1}})\propto\int_{%
\theta_t}P(\theta_{t+1},Y_{t+1}|\theta_t,\mathbf{Y_t})P(\theta_t|\mathbf{Y_t})d\theta_t
\end{equation*}
given before, we have
\begin{align*}
P(\theta_{t+1}|\mathbf{Y_{t+1}})& \propto\int_{\theta_t}P(%
\theta_{t+1}|\theta_t,Y_{t+1})P(Y_{t+1}|\theta_t, \mathbf{Y_{t}})\,P(\theta_t|\mathbf{Y_{t}})d%
\theta_t, \tag*{(7.4)}
\end{align*}
where
\begin{equation*}
P(Y_{t+1}|\theta_t, \mathbf{Y_{t}})=\int_{%
\theta_{t+1}} P(Y_{t+1}|\theta_{t+1},\theta_t, \mathbf{Y_{t}}) P(\theta_{t+1}|\theta_t, \mathbf{Y_{t}})
d\theta_{t+1}
\end{equation*}
by law of total probability, by conditioning on $\theta_{t+1}$. Assuming
$\theta_{t+1}$ is independent of $\mathbf{Y_t}$ given
$\theta_t$, and $Y_{t+1}$ is independent of $\theta_t$ and
$\mathbf{Y_t}$ given $\theta_{t+1}$, we have
\begin{align*}
P(Y_{t+1}|\theta_t, \mathbf{Y_{t}})=\int_{%
\theta_{t+1}} P(Y_{t+1}|\theta_{t+1}) P(\theta_{t+1}|\theta_t)
d\theta_{t+1}=P(Y_{t+1}|\theta_t).
\end{align*}
Thus, (7.4) simplifies as:
\begin{equation*}
P(\theta_{t+1}|\mathbf{Y_{t+1}})\propto
\int_{\theta_t}P(\theta_{t+1}|\theta_t,Y_{t+1})P(Y_{t+1}|\theta_{t})P(\theta_t|\mathbf{Y_t})d\theta_t.\tag*{(7.5)}
\end{equation*}
Equation (7.5) parallels (7.2) and is an alternate to it. It encapsulates the "update first-propagate next" protocol. Note that the key difference  between (7.2) and (7.5) pertains to the feature that the former entailed $P(Y_{t+1}|\theta_{t+1})$ whereas the latter entails  
$P(Y_{t+1}|\theta_{t})$. With $Y_{t+1}$ observed as $y_{t+1}$, $P(Y_{t+1}|\theta_{t+1})$ spawns the \textit{filtering likelihood} $\mathcal{L}(\theta_{t+1};\,y_{t+1})$ whereas 
$P(Y_{t+1}|\theta_{t})$ spawns the \textit{smoothing likelihood} $\mathcal{L}(\theta_{t};\,y_{t+1})=P(y_{t+1}|\theta_t)$. An advantage of the smoothing likelihood over the filtering likelihood is that were $y_t$ an outlier but $y_{t+1}$ not, then a consideration of a likelihood based on $y_{t+1}$ would diminish the ill effects of $y_{t}$.

Filtering under the update first-propagate next protocol is an implementation of equation (7.5) via a simulation starting with the generation of $N$ particles $\theta^{(i)}_t$, $i=1,\ldots, N$ from the distribution of $(\theta_t|\mathbf{Y_t})$ and using each $\theta^{(i)}_t$ to specify a smoothing likelihood $\mathcal{L}(\theta^{(i)}_{t};\,y_{t+1})$ and the ensuing importance weights $w^{(i)}_{t+1}\propto\mathcal{L}(\theta^{(i)}_t;y_{t+1})$. Proceeding as above, going from right to left of (7.5) we have 
\begin{equation*}
P(\theta_{t+1}|\mathbf{Y_{t+1}})\approx\sum_{i=1}^NP(\theta^{(i)}_{t+1}|\theta^{(i)}_t,y_{t+1})
w^{(i)}_{t+1},
\end{equation*}
where $P(\theta^{(i)}_{t+1}|\theta^{(i)}_t,y_{t+1})$ is evaluated via Bayes' Law as
$$
P(\theta^{(i)}_{t+1}|\theta^{(i)}_t,Y_{t+1})\propto P(Y_{t+1}|\theta^{(i)}_{t+1})P(\theta^{(i)}_{t+1}|\theta^{(i)}_t),
$$
assuming that $Y_{t+1}$ is independent of $\theta^{(i)}_{t}$ given $\theta^{(i)}_{t+1}$. Thus with $Y_{t+1}$ observed as $y_{t+1}$, we have 
$$
P(\theta^{(i)}_{t+1}|\theta^{(i)}_t,y_{t+1})\propto \mathcal{L}(\theta^{(i)}_{t+1};y_{t+1})P(\theta^{(i)}_{t+1}|\theta^{(i)}_t).
$$

The schematics of Figure 5 illustrates the above operations. Before closing this sub-section, it is appropriate to cite the recent paper by Sukhavasi and Hassibi (2013) which describes the mechanics of filtering when the observation space (as opposed to the state-space) is quantized by particles.
\begin{figure}[!ht]
\begin{center}
\includegraphics[scale=0.75]{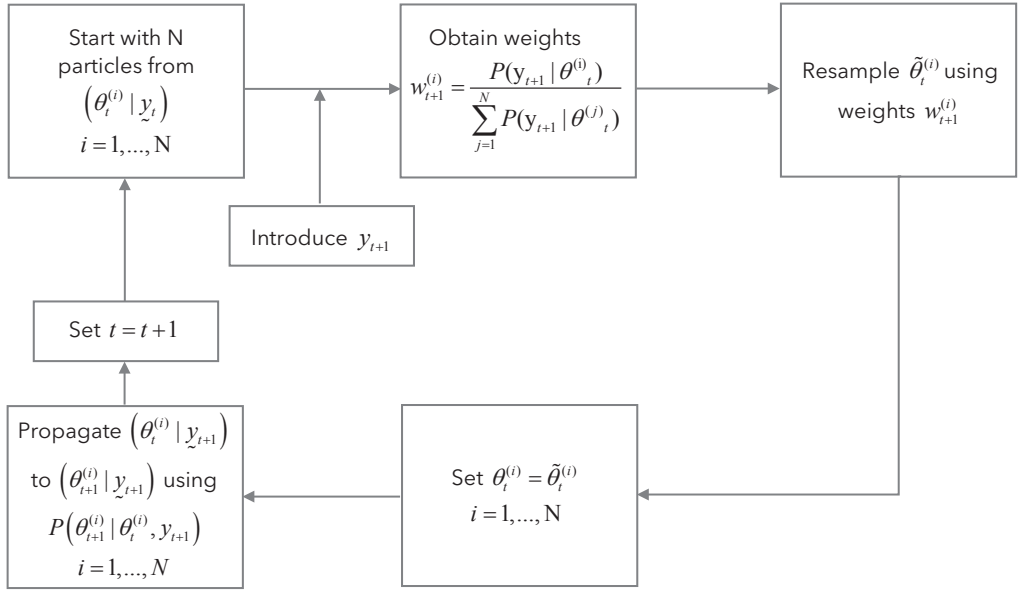}
\end{center}
\label{f5}
\caption{Particle Filtering under Update First$-$propagate Next Protocol.}
\end{figure}
\vspace{0.1in}

\section{Summary, Conclusions, and the Path Forward}

This paper has been primarily written for an audience of applied statisticians, applied probabilists, econometricians, engineers, and time-series analysts, many of whom are familiar with state-space models, but who may not be fully cognizant of the genesis, the evolution and the mathematical underpinnings of such models. The several references citing the work of Mandrekar and his colleagues are given here to provide the reader some sense of what appeals to theoretical probabilists in this arena. Control theorists may find little that is new to them. An exception could be the material of Section 3.1 on the philosophical basis of the Kolmogorov-Wiener setup in the context of quantum theory, and the material of Section 6.2 on the role of the less recognized principle of conditionalization on the  commonly used results in filtering. An adherence to this principle
is philosophically not mandatory, and when not adhered to, it could fundamentally change the nature of some well known results and the algorithms which produce these.

Besides the philosophical material of Sections 3.1 and 6.2, what distinguishes this paper from other
surveys and reviews on filtering is its encompassiveness. Rather than focussing solely on computational or simulation issues, the paper gravitates towards the underlying ideas, and  
traces the key mileposts of the subject which constitute the core of its foundations. 
See Figure 6 whose title is inspired by term "the quark jungle of particle physics."
It starts with the work of Gauss, who laid out a general paradigm for all that is to follow, and then moves on to that of Kolmogorov who put forth a mathematical framework to operationalize Gauss' paradigm. Wiener enters the picture, presumably independently of Kolmogorov, and ends up adding some structure to Kolmogorov's very general setup. But this was not enough; ease of implementation continued to be a problem. This first motivated North to propose the "matched filter," which in turn was followed up by Bode and Shannon, and Zadeh and Ragazzini to push the envelope further by adding structure to Wiener's setup, so that now Kolmogorov's setup had an enhancement in two tiers, the first due to Wiener, and the second due to North, Bode-Shannon, and Zadeh-Ragazzini. These have paved the path towards development of hidden Markov and state-space models. Whereas Shannon and Zadeh have been acknowledged as the originators of information and fuzzy set theory, respectively, the signal role played by them in the development of state-space models warrants a more emphatic recognition. The dates shown in Figure 6 are accurate to the best of our knowledge.

\begin{figure}[!ht]
\begin{center}
\includegraphics[scale=0.6]{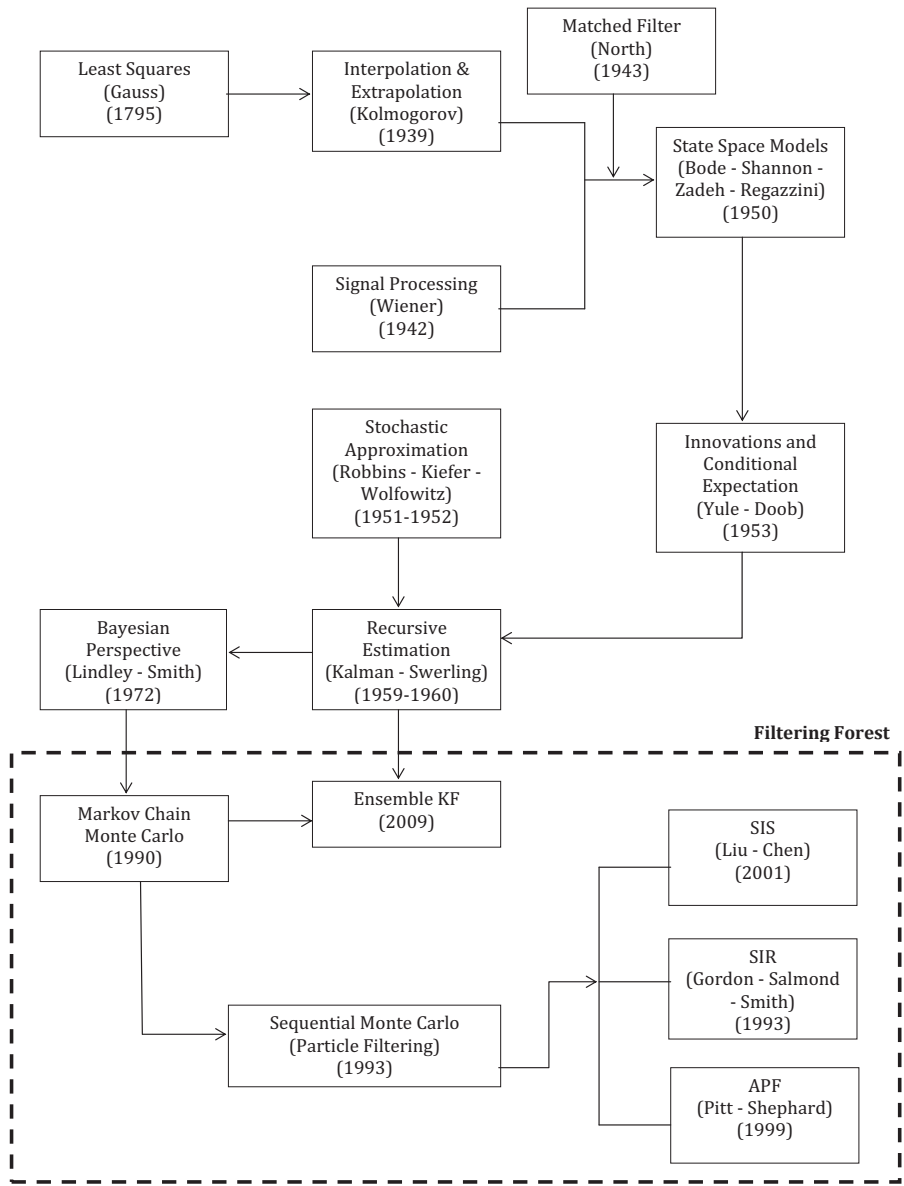}
\end{center}
\label{f6}
\caption{The Journey: From Least Squares to the Filtering Forest.}
\end{figure}
\vspace{0.15in}

Not to be forgotten is the role of statisticians like Robbins, Kiefer, and Wolfowitz in the enhancement and development of state-space models. Noteworthy is the landmark paper of Lindley and Smith (1972), who gave Kalman's algorithm a Bayesian prior to posterior interpretation, and in so doing opened the floodgate for statisticians to join the party. This has been a fortuitous development, because statisticians and applied probabilists have developed powerful computational and simulation tools that have advanced the state of the art of filtering by increasing its efficiency. In exchange, state-space models and filtering techniques enhanced the scope of regression models by making them dynamic, and have enhanced the scope of statistical modeling vis a vis graphical modeling and causal analysis. A recent paper by Smith and Freeman (2011) provides a striking perspective on this reciprocal relationship.

Traditionally, state-space models have primarily been used in signal processing, image analysis, target tracking, astronomical studies, and time series analysis. The era of big data has opened the door to other applications as well, and this is what we mean by path forward. An inkling of this possibility is the work of Li, Holland, and Meeker (2010), which pertains to a problem in reliability. Big data tends to be high dimensional because it is often generated by an array of sensors that are spatially placed and which generate volumes of information in real time. In the application by Li et al. (2010) filtering is done in three dimensions via a matched filter, and the challenge of doing so is addressed by the Fast Fourier Transform. It has often been claimed that the future of reliability and maintainability will be driven by an ability to anticipate failure and to take timely preventive measures by the real time tracking of degradation and wear; see, for example, Lu and Meeker (1993). Sentiments such as these have spawned efforts such as those by Qian and Yan (2015) for using the particle filter to predict useful life of bearings, by Wang, Miao, Zhou, and Zhow (2015) for gear, by Sun, Zou, Wang, and Pecht (2012) for gas turbines, and by Zio and Peloni (2011) for tracking fatigue crack growth. An overview of prognostics based on particle filter methods is given by Jouin, Gouriveau, Hissel, Pera and Zerhouni (2016). More recently, with the advent of self driving cars and airplanes such as the "Dreamliner", filtering techniques have been used to predict the residual lifetimes of rechargeable batteries. Here, degradation is often described by a Brownian motion process with an adaptive drift, and is tracked by a particle filter; see, for example, Wang, Carr, Xu, and Kobbacy (2011), Dalal, Ma, and He (2011), Xing, Ma, Tsui, and Pecht (2013), and Si (2015).

With big data, one  may also need to engage with stochastic processes in high dimensions. The theoretical foundation for doing the above was initiated by Wiener and Masani (1957,1958), and by Masani (1960). The material there is technically demanding, and is mentioned here mainly for sake of historical completeness.
\\

\subsection*{ACKNOWLEDGEMENTS}
The authors thank the editors and referees for their thorough
and constructive reviews which enhanced the scope and quality of the paper. 

One of the authors (NS) was greatly inspired by Professor Thomas Kailath's
(1991) review paper cited below and benefited greatly from its comprehensive
and exhaustive coverage. He thanks Tom for this contribution to his learning, and also his comments on an earlier draft of this paper.
He also thanks Professor William Meeker who made several comments pertaining to scope, coverage, and projecting to the future, and for raising awareness about "matched filters." Finally, Viyadhar (Atma) Mandrekar took many pains alerting him to the more recent mathematical work in filtering.

Nozer Singpurwalla's work was supported in part by the US Office of Naval
Research - Global Grant N62909-15-1-2026 with The City University of Hong
Kong, and The City University of Hong Kong Project Numbers 9380068 and
9042083.

\begin{singlespacing}

\end{singlespacing}

\end{document}